\documentclass{article}
\usepackage[titletoc,title,toc]{appendix}
\usepackage{graphicx}  
\usepackage{amsmath}   
\usepackage[compress]{cite}
\usepackage{amssymb}   
\usepackage{bm} 
\usepackage{dcolumn}
\usepackage{color}
\usepackage[caption=false]{subfig}
\usepackage{mathrsfs}
\usepackage{amsfonts}
\usepackage{longtable}
\usepackage{varioref}
\usepackage{float}
\RequirePackage[colorlinks,citecolor=blue,urlcolor=magenta,linkcolor=blue]{hyperref}
\def\gr{GR}

\labelformat{section}{Section #1} 
\labelformat{subsection}{Section #1} 
\labelformat{subsubsection}{Section #1}
\labelformat{subsubsubsection}{Section #1}
\labelformat{equation}{Eq.~(#1)} 
\labelformat{figure}{Fig.~#1} 
\labelformat{subfigure}{Fig.~\thefigure#1} 
\labelformat{table}{Table~#1} 
\labelformat{appendix}{Appendix #1}
\title{Constraining extra-spatial dimensions with observations of GW170817}

\author{Kabir Chakravarti\footnote{kabir@iucaa.in}~$^{1}$, 
Sumanta Chakraborty\footnote{sumantac.physics@gmail.com}~$^{2,3}$, 
Khun Sang Phukon\footnote{khunsang@iitk.ac.in}~$^{4}$\\
Sukanta Bose\footnote{sukanta@iucaa.in}~$^{1,5}$ and
Soumitra SenGupta\footnote{tpssg@iacs.res.in}~$^{3}$\\
{$^{1}$\small{IUCAA, Post Bag 4, Ganeshkhind, Pune University Campus, Pune 411007, India}}
\\
{$^{2}$\small{School of Mathematical and Computational Sciences}}
\\
{\small{Indian Association for the Cultivation of Science, Kolkata 700032, India}}
\\
{$^{3}$\small{School of Physical Sciences, Indian Association for the Cultivation of Science, Kolkata 700032, India}}
\\
{$^{4}$\small{Indian Institute of Technology, Kanpur, Kanpur 208016, India}}
\\
{$^{5}$\small{Department of Physics and Astronomy, Washington State University, 1245 Webster, Pullman, WA 99164-2814, USA}}}
\begin{document}

\maketitle
\begin{abstract}
We derive the modifications introduced by extra-spatial dimensions beyond the four dimensional spacetime on the macroscopic properties of neutron stars, which in turn affect the gravitational wave spectrum of their binaries. It turns out that the mass-radius relation of the neutron stars, and their tidal deformability, are affected non-trivially by the presence of extra dimensions, and can be used to constrain parameters associated with those dimensions. Implications for I-Love-Q universality relations are also discussed and utilized to obtain a constraint on one such parameter. Importantly, we show, for the first time, that measurements of the component masses and tidal deformabilities of the binary neutron star system GW170817, constrain the brane tension in the single brane-world model of Randall and Sundrum to be greater than $35.1~\textrm{GeV}^{4}$. This work opens up the possibility of making such a constraint more robust by improving the modelling of binaries on the brane in the future.
\end{abstract}
\section{Introduction}

The recent success of Gravitational Wave (GW) detectors in observing mergers of binary Black Holes (BHs) and binary Neutron Stars (NSs) has opened up a plethora of possibilities for testing and understanding fundamental physics \cite{Abbott:2017vtc,TheLIGOScientific:2016src,Abbott:2016blz,TheLIGOScientific:2017qsa}. With GW astronomy getting established on a firm observational footing, it is now possible to comprehend and, better still, probe fundamental questions of nature. In this article we explore one such question, namely, whether there exist extra spatial dimensions beyond the four-dimensional spacetime that we most readily access. We do so by capitalizing on the recent GW observation of the binary NS merger GW170817 \cite{TheLIGOScientific:2017qsa}. One limitation of our study is that we restrict ourselves to non-spinning stars. This may not be a bad approximation since Galactic double NS systems suggest a dimensionless spin parameter of less than 0.05 for binary NSs in the LIGO-Virgo band~\cite{Lorimer:2008se}. Nonetheless we aim to revisit this analysis for spinning stars in the future.

Gravitational-wave data, including that of GW170817, are typically analyzed from the perspective of a four-dimensional observer in General Relativity (GR). However, deviations from GR have been constrained, e.g., in Ref.~\cite{Abbott:2018lct,LIGOScientific:2019fpa}. Here we inquire whether the observation of GW170817 allows any room for such deviations, specifically, in the realm of an effective four-dimensional theory that originates from a higher dimensional theory of gravity \cite{Randall:1999ee,Randall:1999vf,Chamblin:1999by,Shiromizu:1999wj}. Originally, extra dimensions were introduced in attempts at unifying gravity and electrodynamics by Kaluza and Klein, but later appeared quite naturally in string theory, which requires ten or more dimensions~\cite{Green:1987sp}. More recently, extra dimensions have resurfaced with the potential to solve the gauge hierarchy problem~\cite{ArkaniHamed:1998rs,Antoniadis:1998ig,Rubakov:1983bb,Csaki:2004ay}. The latter addresses the fact that fundamental energy scales in physics appear uncorrelated and hierarchical in nature: The energy scale of electro-weak symmetry breaking ($\sim 10^{3}~\textrm{GeV}$) is apparently completely disconnected from the Planck scale ($\sim 10^{18}~\textrm{GeV}$). This in turn requires an unnecessarily large fine tuning in order to get the observed mass of the Higgs Boson at the Large Hadron Collider (LHC) \cite{Csaki:2004ay}. 

We model the extra-spatial dimension as one additional spacelike dimension; the effective gravitational field equations are derived by imposing orbifold symmetry on that dimension. As expected, these equations inherit additional corrections from the bulk geometry~\cite{Shiromizu:1999wj,Dadhich:2000am,Maartens:2003tw}. We use the properties of GW170817 to subject those corrections to strong-field tests of gravity. The presence of extra dimension(s) would leave very specific signatures on the GWs emitted by merging BHs or NSs and will affect all three phases of the signal, namely the inspiral, the merger and the post-merger phase, most notably the ringdown phase. The ringdown phase is interesting as well as important in its own right, since not only it can be understood analytically, but also it provides the fundamental modes of BHs, known as the Quasi Normal Modes (QNMs)~\cite{Chandrasekhar:1985kt,Nollert:1999ji,Kokkotas:1999bd,Berti:2009kk,Konoplya:2011qq}. For BHs in the effective four-dimensional gravitational theory as well, the QNMs carry distinct signatures of the presence of extra dimensions~\cite{Toshmatov:2016bsb,Chakraborty:2017qve}. Even though it is possible to write down such an effective description for theories beyond \gr\ as well (see e.g., \cite{Chakraborty:2014xla,Chakraborty:2015taq,Chakraborty:2015bja} for theoretical analysis and \cite{Bhattacharya:2016naa,Mukherjee:2017fqz,Banerjee:2017hzw} for observational probes), here we will content ourselves with the general relativistic situation alone and using the binary neutron star merger event, would like to put the theory to the strong field test of gravitational interaction.

We utilize the GW170817 bounds on the quadrupolar deformation of a NS in its binary companion's tidal field \cite{Lattimer:2004pg,Hinderer:2007mb,Flanagan:2007ix,Binnington:2009bb,Damour:2009vw} to constrain an important parameter of Brane-world models, namely, the brane tension $\lambda_{\rm b}$, which counters the influence of the negative cosmological constant of the (anti-de Sitter) bulk on the brane~\cite{Dadhich:2000am,Maartens:2003tw}. One can indeed analyze the modifications due to the presence of higher dimensions to NS tidal deformations \cite{Chakravarti:2018vlt} caused by a companion in a binary; for some recent works regarding modifications of tidal deformability parameter in other theories beyond \gr, see Refs. \cite{Cardoso:2017cfl,Yazadjiev:2018xxk,Olmo:2019flu}). In particular, it was demonstrated in \cite{Chakravarti:2018vlt} that the tidal deformability parameter for BHs in the presence of extra dimension is \emph{non-zero} and, in fact, \emph{negative}, while for NSs with the same central density it is smaller than the corresponding value in \gr. Following this result we would like to understand the modifications arising in the tidal deformation of NSs due to the presence of extra dimensions. Any possible change can be used to test the universality of the I-Love-Q relation~\cite{Yagi:2013bca} as well as the mass-radius relation of NSs in the presence of extra dimensions. We will show here that both these relations, and the electromagnetic observation of GW170817, will constrain the parameter $\lambda_{\rm b}$ in a novel way that is consistent with sub-millimeter tests of gravity \cite{Long:2002wn}, but is likely affected by different systematics. Moreover, it must be emphasized that this is a first attempt to constrain extra-spatial dimensions, in the context of single brane model of Randall-Sundrum, through the gravitational wave observations. This analysis may be improved further by more realistic modelling of binaries and by combining more observations as and when they become available in the future.

The paper is organized as follows: We start by introducing the higher dimensional model under consideration along with the basic equations governing gravitational dynamics in \ref{Sec_extra_dim}. The mass-radius relation originating from the modified gravitational field equations has been presented in \ref{Sec_mass_radius}. Following which, in \ref{Sec_universal}, we have presented the universality of the I-Love-Q relations. Finally the implications of the GW170817 event for the presence of higher dimensions have been described in \ref{Sec_GW170817}, before concluding in \ref{Sec_Conclusion}. The details of the gravitational field equations within the neutron star have been presented in \ref{AppA}.

\textit{Notation and convention:} We set $\hbar=1=c$ in all the mathematical expressions. Greek indices are used to denote four-dimensional quantities and the mostly positive signature convention is followed throughout the article. 
\section{Spacetime with extra dimension: A brief introduction}\label{Sec_extra_dim}
 
In this section, we will briefly review the setup with which we are working in this paper. Among the various models of higher dimensions, those proposed by Randall and Sundrum are of primary significance \cite{Randall:1999ee,Randall:1999vf}. There are two variants of such models, one involving compactified extra dimensions, leading to a tower of massive Kaluza-Klein modes, while the other one corresponds to large extra dimensions. In the case of merger of binary NSs, the energy emitted is not significant enough to excite the massive Kaluza-Klein modes and hence we will be concentrating on the model with large extra dimension (for a review, see \cite{Maartens:2003tw}). 

In such a model with large extra dimensions, it is instructive to work with an effective gravitational theory in the four-dimensional spacetime, known as the brane, that has been inherited from a five-dimensional spacetime -- the bulk. Such an effective theory can be obtained by projecting the Einstein's equations from the bulk onto the brane. This yields the following equations governing the dynamics of gravity on the brane~\cite{Maartens:2003tw,Chakravarti:2018vlt}:
\begin{align}\label{Eff_Eq}
~^{(4)}G_{\mu \nu}+E_{\mu \nu}=8\pi G\left\{T_{\mu \nu}+\frac{6}{\lambda_{\rm b}}\Pi_{\mu \nu}\right\} \,,
\end{align}
where $G$ is Newton's gravitational constant, $E_{\mu \nu}$ is the projected bulk Weyl tensor on the brane, $T_{\mu \nu}$ is the brane energy-momentum tensor, $\Pi_{\mu \nu}$ is a second-rank tensor constructed from various quadratic combinations of $T_{\mu \nu}$, and $\lambda_{\rm b}$ is the brane tension \cite{Shiromizu:1999wj,Maartens:2003tw}. The GR limit is achieved by taking $\lambda _{b}\rightarrow \infty$, where $E_{\mu \nu}$ also vanishes \cite{Maartens:2003tw}.

Since the equilibrium structure of any compact object on the brane is governed by the above equation, the interior of any NS will also follow the same. Thus, it will inherit two additional contributions, over and above \gr, namely non-local corrections from the bulk through $E_{\mu \nu}$ and quadratic corrections to the matter energy momentum tensor through $\Pi_{\mu \nu}$. These contributions must be matched at the surface of the NS with its exterior, where $\Pi_{\mu \nu}$, along with $T_{\mu \nu}$, identically vanishes. The matching condition, at the surface of the star, takes the following form \cite{Germani:2001du,Chakravarti:2018vlt},
\begin{align}\label{TLN_v3_07}
p_{\rm eff}(R)+U_{\rm int}(R)+2P_{\rm int}(R)=U_{\rm ext}(R)+2P_{\rm ext}(R)
\end{align}
where $R$ is the radius of the NS and we have decomposed the bulk Weyl stress $E_{\mu \nu}$ as, $E^{\mu}_{\nu}\propto \textrm{diag}(-3U,U+2P,U-P,U-P)$. Thus in full generality, the Tolman-Oppenheimer-Volkoff (TOV) equation in the interior of the star must be treated as a two fluid model. This makes the equations for gravitational perturbation in the interior of the star non-separable and hence computation of the tidal Love number becomes complicated. As evident from \ref{TLN_v3_07}, the most economic way to get  around this problem is to assume that $U_{\rm int}(R)=0=P_{\rm int}(R)$, i.e., the effects from extra dimension due to the Weyl stress identically vanishes in the interior of the NS. This, besides simplifying the analysis, also reduces the equations representing gravitational perturbations in the interior of the NS to a single fluid system. Thus we would like to emphasize that for equilibrium, \emph{it is absolutely essential that $E_{\mu \nu}$ is non-zero in the region exterior to the surface of the star}. However, in the interior of the star we can safely argue for vanishing contribution from the bulk Weyl tensor.

Also note that in the context of binary NS system, the boundary condition must be imposed on the surface of both the NSs. Since the exterior bulk Weyl tensor is identical for both, the radii of the NSs will be determined by the radial profile of the internal pressure, whenever \ref{TLN_v3_07} is satisfied. This is identical to the case of \gr, where also the exterior solution is almost identical (at least in the linear regime) and the radius of the star is determined by the vanishing of pressure. In this case, the effective pressure does not vanish, but gets matched with the bulk Weyl stress in the exterior. Hence binary systems can also be described with vanishing Weyl stress in the interior. Hence for our purpose, it will suffice to work with vanishing bulk Weyl stress in the interior of the NS, such that $\Pi_{\mu \nu}$ at the surface of the NS is balanced by a non-zero $E_{\mu \nu}$ in the exterior of the star (for a similar treatment of compact objects in the brane world, see, \cite{Germani:2001du,Wiseman:2001xt,Visser:2002vg,Ovalle:2014uwa,Herrera-Aguilar:2015koa,Felipe:2016lvp,HeydariFard:2009tj,Chakravarti:2018vlt}).

With this setup it is straightforward to determine the master equation governing the even-parity gravitational perturbations \cite{Chakravarti:2018vlt}. The main departure of the perturbation equation from \gr\ is through the presence of effective energy density $\rho_{\rm eff}$ and isotropic pressure $p_{\rm eff}$, which are related to the energy density $\rho$ and pressure $p$ of the brane matter as
\begin{align}\label{Eq_rho_p}
\rho_{\rm eff}=\rho\left(1+\frac{\rho}{2\lambda_{\rm b}}\right);\qquad p_{\rm eff}=p+\frac{\rho}{2\lambda _{\rm b}}\left(\rho+2p\right)~.
\end{align}
The ensuing modification in the Equation of State (EoS) parameters~\cite{Chakravarti:2018vlt} of the NS owing to the presence of the extra dimension can reveal itself in various ways. First, the tidal Love number will be modified, which will change the stiffness of the effective EoS parameter~\cite{Chakravarti:2018vlt}. Second, the mass-radius relation for NSs will pick up  corrections for finite $\lambda _{\rm b}$. Lastly, the universality relation between moment of inertia, tidal Love number and quadrupole moment of NSs may be affected. We  explore all these possibilities in the light of GW170817.
\section{Mass-radius relation on the brane}\label{Sec_mass_radius}
 
In this section, we will construct the relation between the mass of a NS with its radius, for various possible choices of the matter EoS and different choices of the brane tension parameter $\lambda_{\rm b}$. Determining such relations will require solving the full set of TOV equations originating from the effective gravitational field equations, expressed by \ref{Eff_Eq} and conservation of matter energy momentum tensor. These equations are written down in \ref{AppA} and have been solved using available numerical techniques in order to determine the associated mass-radius relation.

\begin{figure}[h]
\centering
\includegraphics[scale=0.45]{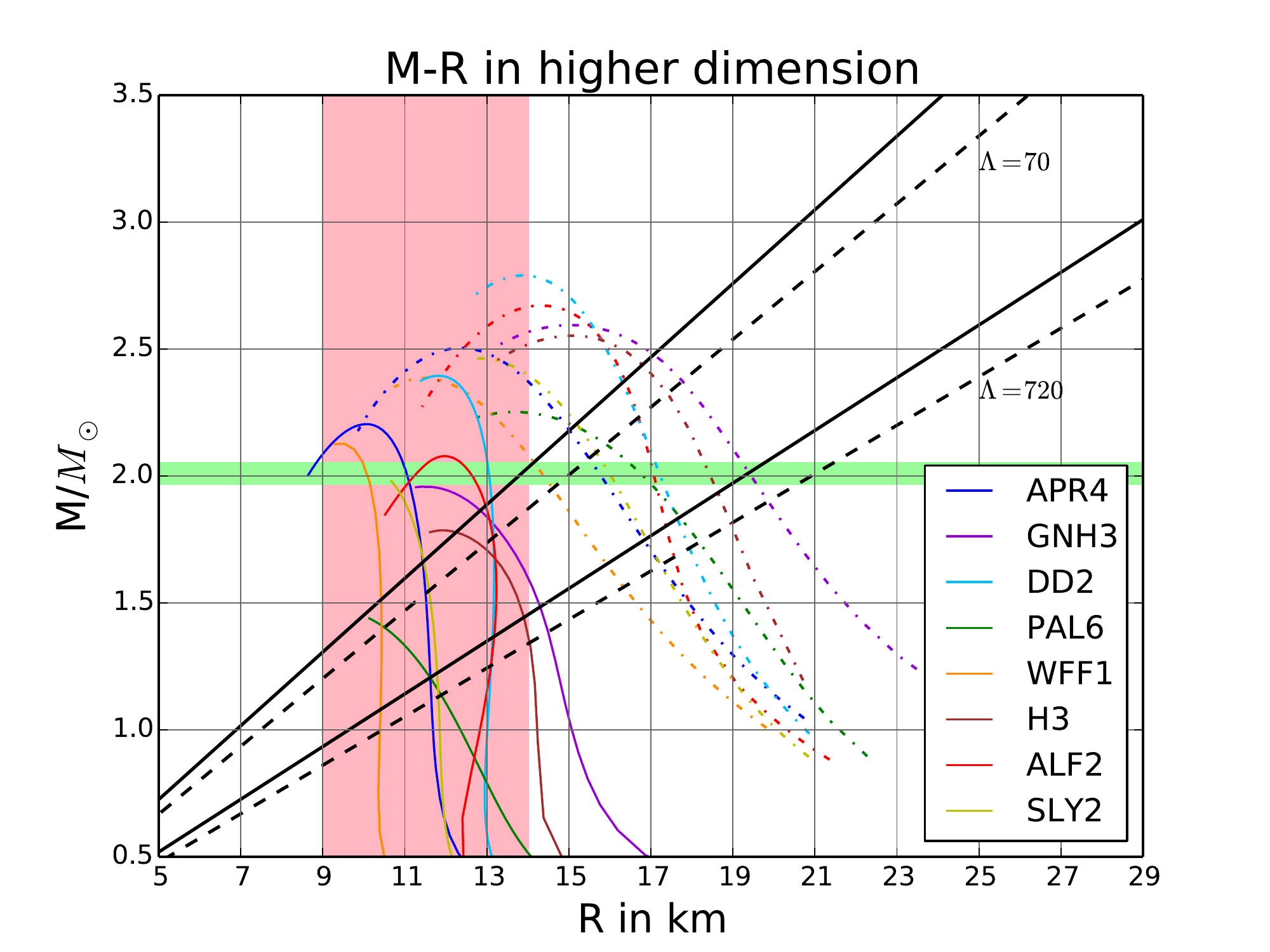}
\caption{Mass-radius curves for a handful of EoSs in GR (solid) and brane world model with $\lambda_{\rm b} = 35.1~{\rm GeV}^{4}$ (dotted) are presented. Also shown are the $\Lambda=70$ and $\Lambda=720$ lines for both, where $\Lambda$ is the tidal Love number. The $\sim 2M_{\odot}$ lower bound on the maximum mass of a NS is shown in green and the GW170817 bounds on the radius of the NSs as a vertical pink strip. As evident, all $\lambda _{\rm b}\leq 35.1~\textrm{GeV}^{4}$ values are ruled out as they do not fall in the range of masses that galactic double NSs and the GW170817 NSs occupy.}
\label{Fig_M-R}
\end{figure}

Following which we have constructed the mass ($M$) - radius ($R$) curves for NSs with several different EoSs, both in the absence and presence of the extra dimension, and presented them in \ref{Fig_M-R}. Particularly, note that for a given radius, the mass of a NS increases with decreasing brane tension, which follows from \ref{Eq_rho_p}. This suggests that introduction of extra dimensions makes the star more compact for the same central density. Further, \ref{Fig_M-R} indicates a possible way to constrain the brane tension, namely, by utilizing observed upper bounds on the mass and radius of NSs. The lower bound on the maximum mass of a stable NS that any EoS must produce is $\sim 2M_{\odot}$ \cite{Demorest:2010bx,Antoniadis:2013pzd}\footnote{Even though the analysis of \cite{Demorest:2010bx,Antoniadis:2013pzd} assumes \gr, the change due to a non-zero $\lambda_{\rm b}$ on the mass measurement has negligible effect on our results.} (the green horizontal strip) and the radius of NS with mass between 1 and 2$M_\odot$ is expected to be in the range $9~\textrm{km}$ to $14~\textrm{km}$ \cite{Lattimer:2015nhk} (the pink vertical strip). In this context, it is worth emphasizing that, there are some recent results \cite{Linares:2018ppq,Cromartie:2019kug} where the mass of certain NSs were determined to be slightly larger than $2M_{\odot}$. However, the posteriors from the LIGO-Virgo collaboration assumes the $2M_{\odot}$ bound. To be consistent with those, we did not consider here the effect of allowing higher maximum masses for the NSs. We hope to pursue this aspect in a future work. Further, as evident from \ref{Fig_M-R}, all of the EoSs selected here (to cover the aforementioned radius range and respect the minimum mass requirement) in \gr\ (solid lines, and with $\lambda _{\rm b}\rightarrow \infty$) satisfy these constraints, while if $\lambda_{\rm b}\leq 35.1~\textrm{GeV}^{4}$ (dashed lines) all of those EoSs have much larger radii. This effect on the brane tension parameter $\lambda_{\rm b}$ suggests that $M-R$ posteriors from observations of merger of binary NSs can be used to constrain the value for $\lambda_{\rm b}$. However, what are measured with GW observations of binary NSs are the masses and the tidal deformability parameter $\Lambda$. Therefore, to utilize the above idea, we will need to study the effect of $\lambda_{\rm b}$ on $\Lambda$ as well. This is where the study of the universality relation between the tidal Love number and other characteristic parameters of the NS becomes important, which we will work out in the next section.

\begin{figure}[h]
\centering
\includegraphics[scale=0.47]{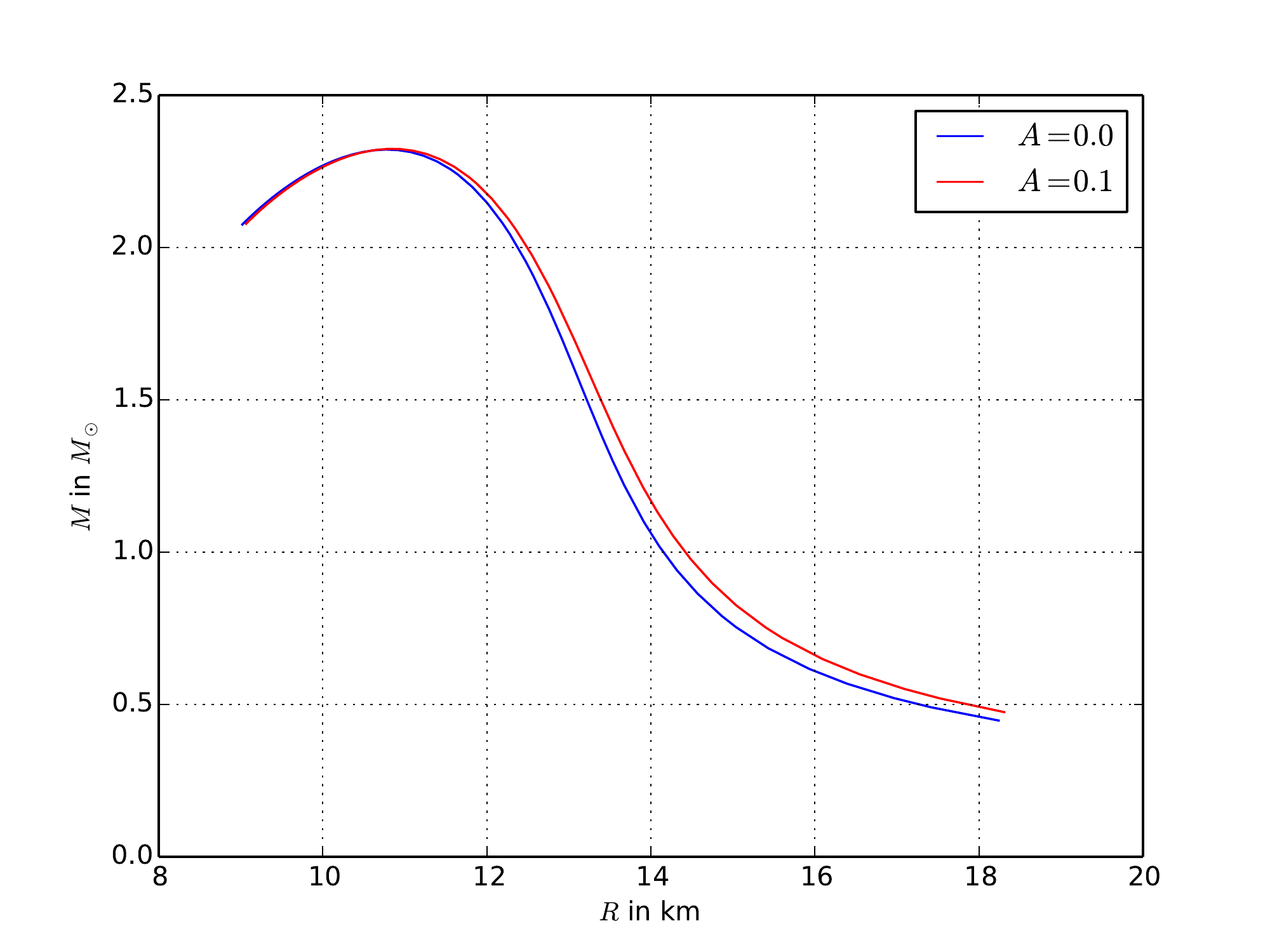}
\caption{The departure of the mass-radius relation in the presence of non-zero Weyl stress in the interior of a NS has been presented. The quantity $A$ corresponds to the dimensionless ratio $\{6\mathcal{U}/\lambda_{\rm b}\rho_{\rm eff}(8\pi G_{4})^{2}\}$. The curve on the left denotes the case we have considered in this work, namely vanishing Weyl stress in the interior (i.e., $\mathcal{U}=0$). On the other hand, the mass-radius relation with non-zero Weyl stress is the one on the right and differs from the situation with vanishing Weyl stress. Hence the mass-radius relation drifts to the right as we include non-zero Weyl stress in the interior of the star. This will further strengthen the bound on $\lambda_{\rm b}$. See text for more discussions.}
\label{Fig_Weyl}
\end{figure}

Finally, for completeness and to see exactly what kind of modifications are brought in by a non-zero bulk Weyl stress in the interior, we have also considered how a non-zero bulk Weyl stress in the interior of the star modifies the mass-radius relation. It turns out that the mass-radius relation gets shifted further towards the right, i.e., it leads to a larger radius at a fixed mass, as the strength of the bulk Weyl tensor increases (see \ref{Fig_Weyl}). Thus with increasing the Weyl stress in the interior, the bound on $\lambda_{\rm b}$ will be further strengthened. Therefore, the bound on $\lambda_{\rm b}$ derived in this work, in the case of vanishing bulk Weyl stress, corresponds to the most conservative estimate of the brane tension. Addition of non-zero Weyl stress will make the value of $\lambda_{\rm b}$ larger than the estimation presented in this work.

As a passing remark, we would like to emphasize that the presence of a non-zero bulk Weyl stress in the interior of the NS will also modify other aspects of the emission process of gravitational wave. As pointed out earlier, the gravitational perturbation equation in the interior will not decouple and hence the tidal Love number associated with deformability of the NS will certainly be different. Since the tidal Love number explicitly appears in the phasing of the gravitational wave signal, there will indeed be modifications in the emission process of gravitational waves from binary NS merger event. However, following \ref{Fig_Weyl}, it is plausible that such corrections, for reasonable bulk Weyl stress, will indeed be smaller.

\section{I-Love-Q with extra dimensions}\label{Sec_universal}

\begin{figure}
\centering
\includegraphics[scale=0.6]{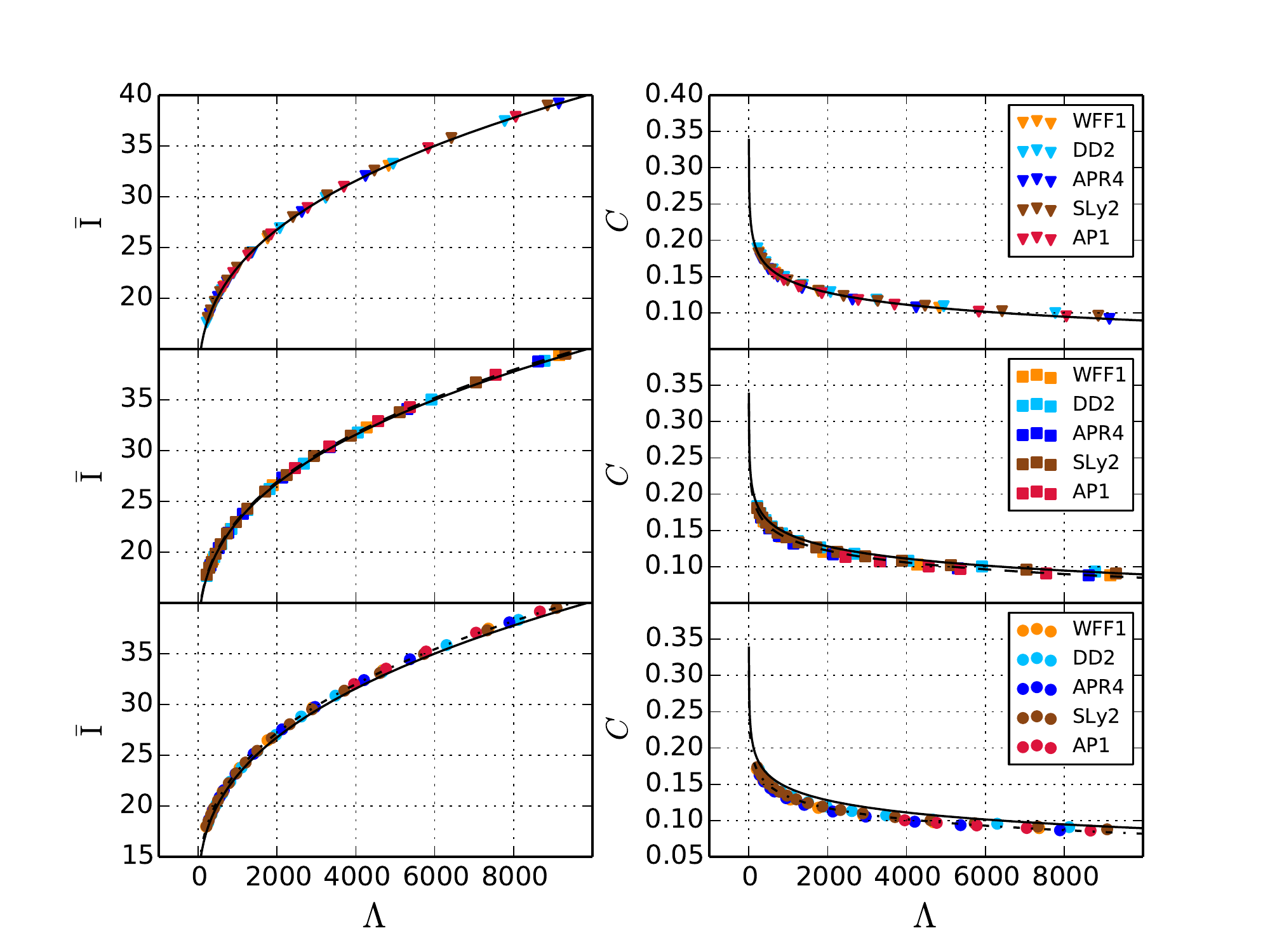}
\caption{The correlations between moment of inertia (I) and compactness (C) of a NS have been depicted with the tidal Love number $\Lambda$, with or without extra dimensions. As evident these correlations depict universal behaviour among these parameters of the NS, irrespective of theoir EoSs. The universality relations for GR are presented in the top row, followed by those for Brane-world models with $\lambda_{\rm b}=1.11\times10^{2}~\textrm{GeV}^{4}$ (middle) and $\lambda_{\rm b}=35.1~\textrm{GeV}^{4}$ (bottom). It is clear from the plots that even though the correlations are universal even in the presence of extra dimensions, they differ from that in GR. See text for more discussions.}
\label{Fig_I_L_Q}
\end{figure}

As solutions of the TOV equations, NSs, have been found to exhibit certain correlations between their Moment of Inertia $I$, Love number $k_{2}$, and their quadrupole moment $Q$, irrespective of the EoS employed. This is the well known I-Love-Q relation, which sometimes is also referred to as the {\em universality relation} of the first kind \cite{Yagi:2013bca,Yagi:2013awa,Yagi:2016bkt,Gupta:2017vsl}. (Note that $\Lambda=(2/3)k_{2}C^{-5}$, where $C=GM/R$ is a measure of the NS compactness.) The effect of a finite $\lambda_{\rm b}$ on such relations is presented in \ref{Fig_I_L_Q}, which shows that they hold true for a few different values of $\lambda_{\rm b}$. More importantly, the universality relation for all choices of $\lambda_{\rm b}$, including $\lambda_{\rm b}\sim 35.1~\textrm{GeV}^{4}$, differs from the universality relation for GR. Thus, even in the presence of extra dimensions, the I-Love-Q relation \emph{remains universal but differs in its parametrization} from the one in GR. We find the general form of the universality relation between $C$ and $\Lambda$ in presence of extra dimensions to be:
\begin{align}\label{univall}
C= \Lambda^\alpha + C_0~,
\end{align}
where $\alpha$ and $C_0$ are constants varying with the brane tension $\lambda_{\rm b}$. Possible estimations of these two constants with the brane tension are presented in \ref{Table_01}.
\begin{table}
\centering
\caption{Best fit values of the parameters $\alpha$ and $C_{0}$ associated with the universality relation \ref{univall} are presented here for different choices of the brane tension.}
\begin{tabular}{cccccc} 
\hline 
$\lambda _{\rm b}$ (in $\textrm{GeV}^{4}$) & $\alpha$       &  $C_{0}$  \\ 
\hline\hline 
$\infty (\textrm{GR})$                                  &  $-0.0312$    & $-0.6607$ \\
$1.11\times 10^{3}$                                                         &  $-0.0315$    & $-0.6597$ \\
$1.11\times10^{2}$                                                          &  $-0.0301$    & $-0.6727$ \\
$4.42\times10^{1}$                                                          &  $-0.0283$    & $-0.6879$ \\
$3.51\times10^{1}$                                                          &  $-0.0286$    & $-0.6893$  \\
\hline 
\end{tabular}
\label{Table_01}
\end{table}

This scenario, apart from highlighting yet another observational avenue to understand extra dimensions, also sheds some light on the origin of the universality relations themselves. As argued in \cite{Yagi:2013bca,Yagi:2013awa,Yagi:2016bkt}, such universality relation possibly originates from the fact that all the EoSs behave in an identical manner in the outer layers of the star, where most of the contribution to $Q$ and tidal deformability arises, or is a result of a BH being the limiting configuration of a NS and, therefore, may be related to the no-hair theorem. We note that if the second scenario is true then the universal behaviour must hold even for non-zero $\lambda_{\rm b}$. Intriguingly, as evident from \ref{Fig_I_L_Q} that is exactly what happens in the presence of extra dimensions, namely that the universality relation holds good at $\lambda_{\rm b}=35.1~\textrm{GeV}^{4}$ but differs from GR. Given the above universality relation \emph{even} in the presence of extra dimensions, we will demonstrate in the next section, how it can be used to extract relevant information for the binary NS merger event GW170817 and hence constrain the brane tension $\lambda_{\rm b}$.

\begin{figure*}
\centering
\subfloat[GR ($\lambda_{\rm b}=\infty$). \label{Fig_4_a}]{\includegraphics[scale=0.45]{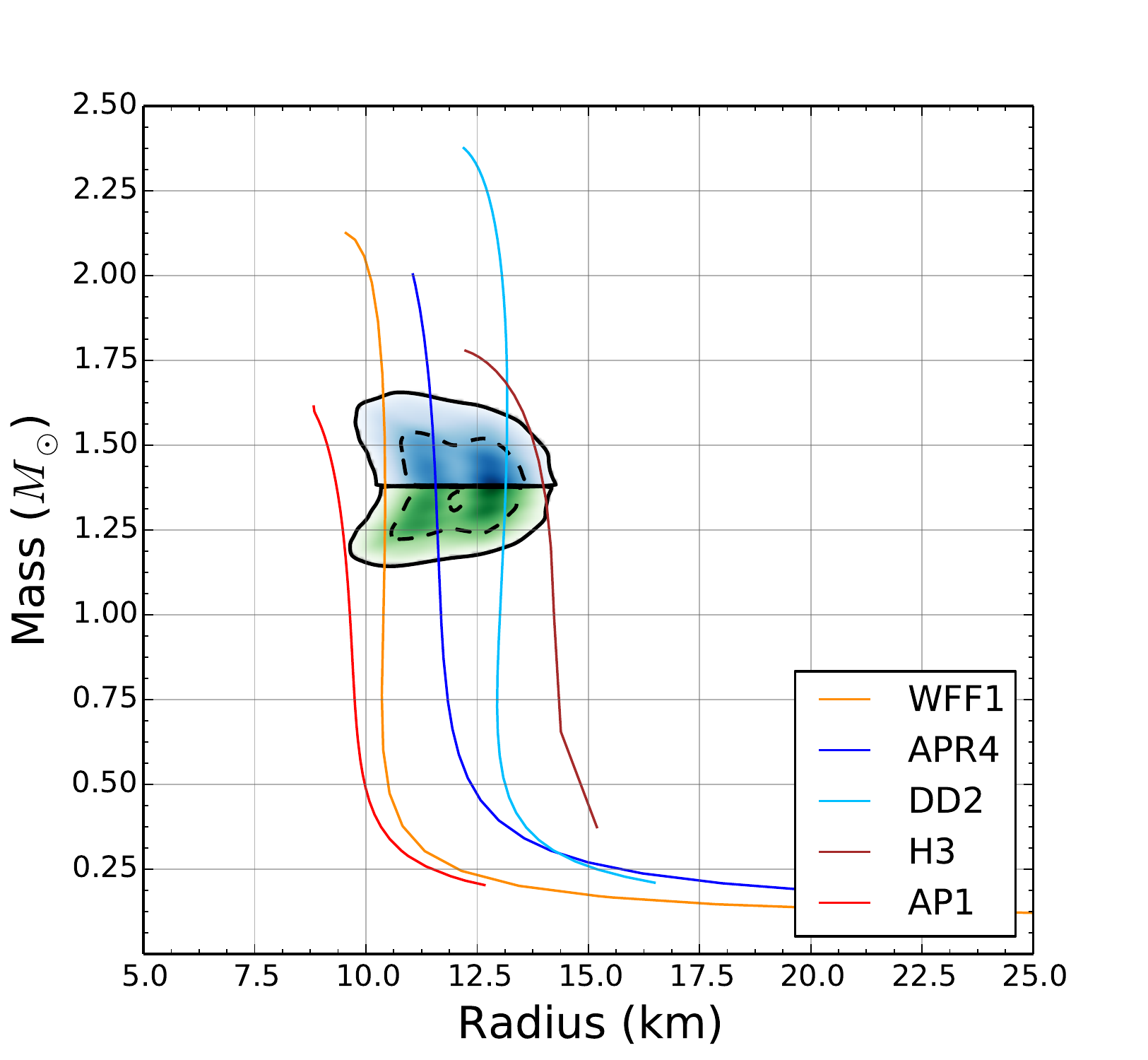}}
\qquad
\subfloat[$\lambda _{\rm b}=44.2~{\rm GeV}^{4}$. \label{Fig_4_b}]{\includegraphics[scale=0.45]{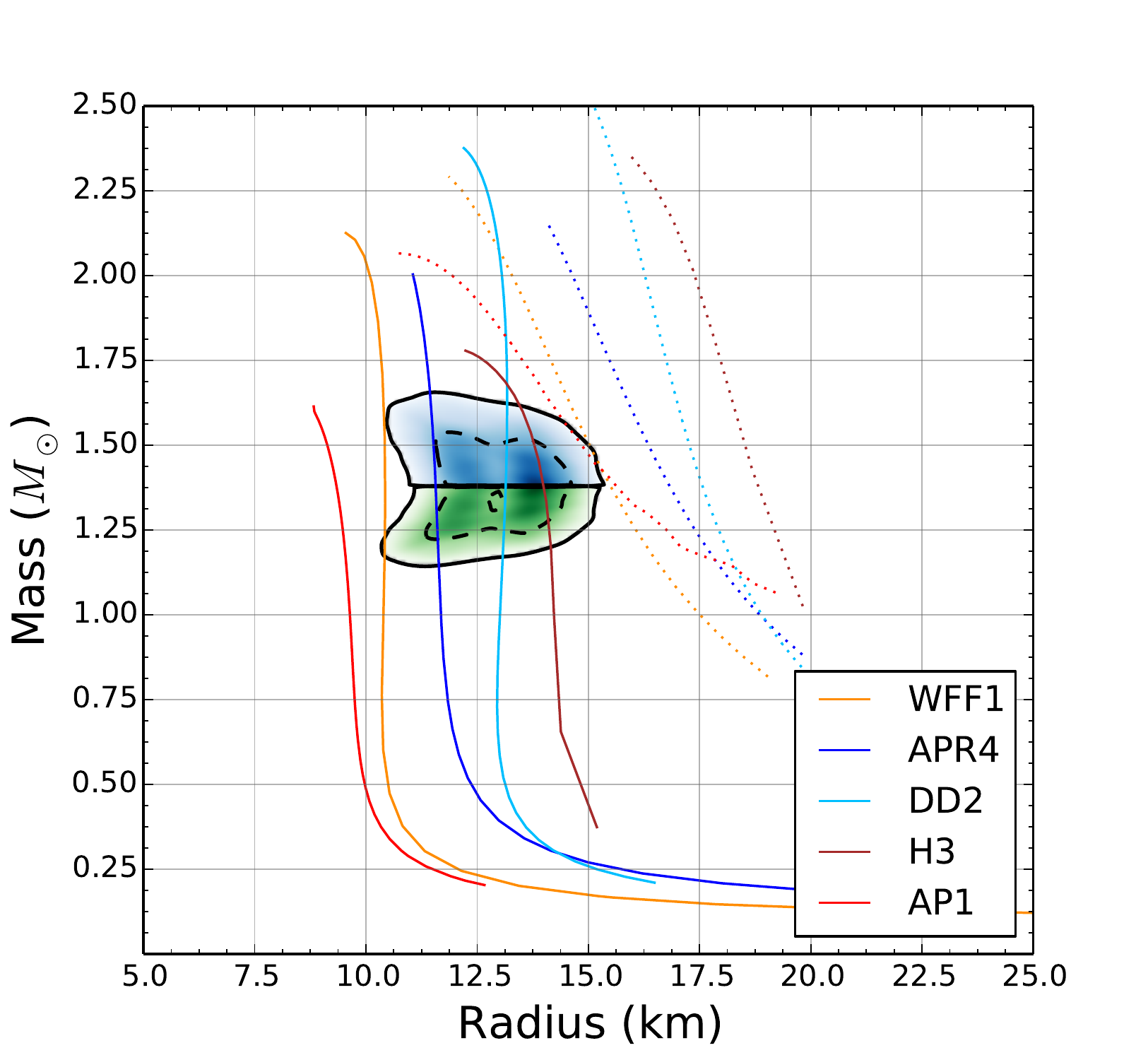}}
\qquad
\subfloat[$\lambda _{\rm b}=35.1~{\rm GeV}^{4}$. \label{Fig_4_c}]{\includegraphics[scale=0.45]{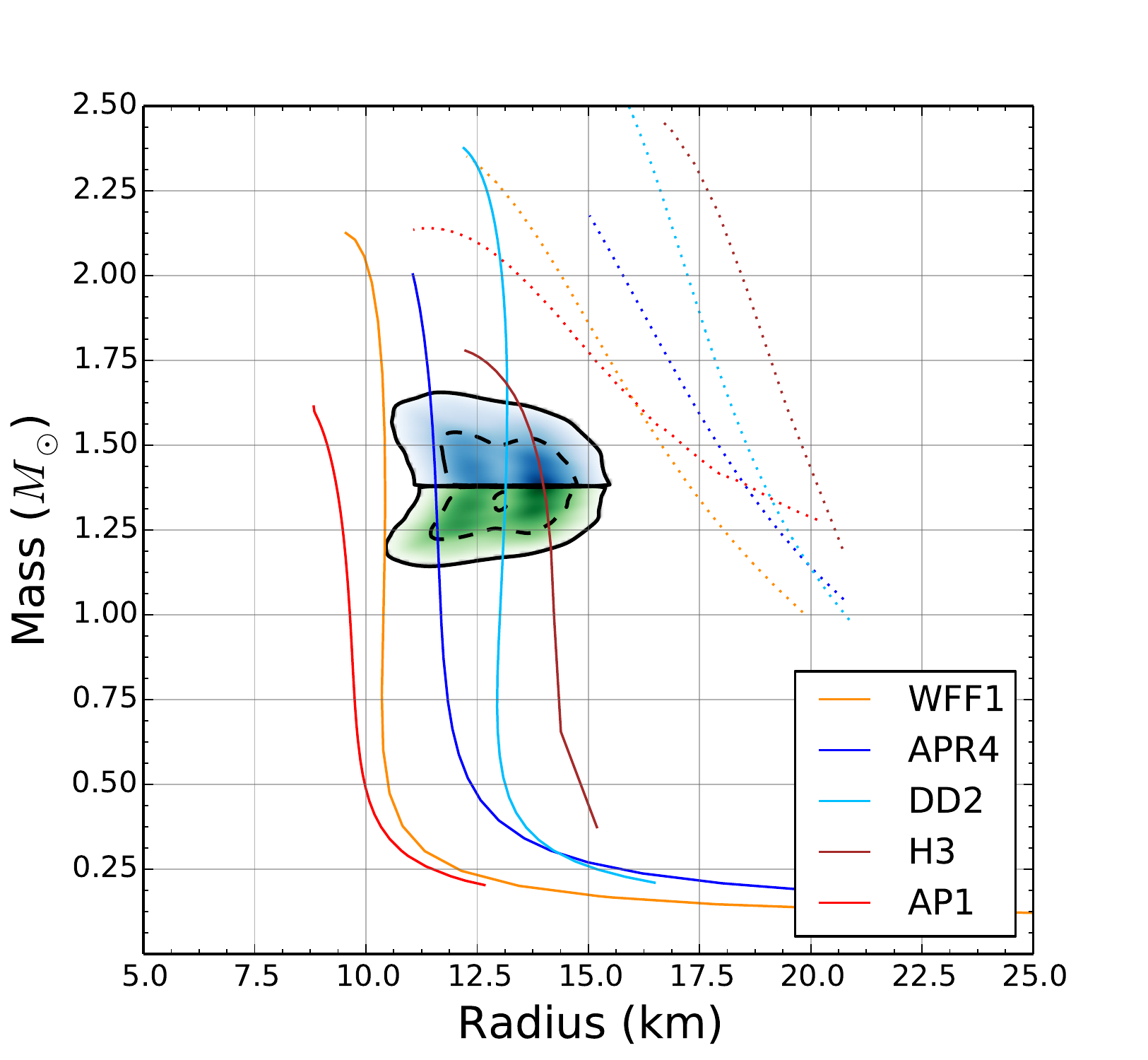}}
\caption{In the plot (a) above, all solid EoS lines denote their respective mass-radius curves for NSs in \gr\ (i.e., $\lambda_{\rm b} \to \infty$). The dotted lines in plots (b) and (c) are for finite values of the brane tension $\lambda_{\rm b}$. These plots explicitly demonstrate that when $\lambda_{\rm b}$ decreases, they move away from the posteriors of GW170817, while the solid curves are retained for reference. The 90\% C.L. posteriors for the primary (blue) and secondary (green) components of GW170817 are shown bounded by solid lines in each figure. (The 50\% credible regions are delineated with dashed lines.) Both the GW170817 posterior as well as the $M-R$ curves of the various EoSs shift to the right as $\lambda_{\rm b}$ decreases. However, the latter shift more than the posterior, which allows us to constrain $\lambda_{\rm b}$ to be greater than $35.1{\rm GeV}^{4}$ since even the softest EoS in this set, AP1, crosses well beyond the $M-R$ region that is allowed by the posterior for that value of $\lambda_{\rm b}$.}
\label{Fig_4}
\end{figure*}
\section{Implications of GW170817}\label{Sec_GW170817}

In this section, we will use the gravitational wave data from the GW170817 measurement to put constraints on the brane tension $\lambda_{\rm b}$. Below we elucidate the steps involved in such an analysis and subsequently demonstrate the result. In order to constrain $\lambda_{\rm b}$, we use its effect on (a) mass-radius curves of NSs and (b) the GW170817 posteriors in the $M-R$ plane. In \ref{Fig_4_a}, all solid EoS lines denote their respective mass-radius curves for NSs in \gr\ (i.e., $\lambda_{\rm b} \to \infty$). The 90\% credible regions for posteriors of the parameters of the primary (blue) and secondary (green) components of GW170817 are shown by solid lines in each plot. (The 50\% credible regions are delineated with dashed lines.) The posteriors shown here are obtained from Ref.~\cite{Vallisneri:2014vxa}, which are also shown in Ref.~\cite{Abbott:2018exr} as those for the parameterized EoS where a lower limit on the maximum mass of $1.97M_\odot$ is imposed. Note that the universality relations for \gr\ can be used to deduce $\Lambda(M,R)$ of a star at any $(M,R$) point in this figure. Indeed, this $(M,R)$ posterior was constructed from the posterior first obtained in the $(M,\Lambda)$ space, $p(M,\Lambda)$, since the latter pair of parameters are directly measured from the phasing of the GW signal. For this construction, the radius of the star is deduced from $(M,\Lambda)$ by using the GR universality relation. Thus, the posterior in the $M-R$ plane is just $p(M,\Lambda(M,R))$. To obtain the corresponding posteriors for a finite $\lambda_{\rm b}$, we ask what the radius will be of the same $(M,\Lambda)$ star in the corresponding Brane-world model. This gives a somewhat larger value for the radius relative to that in GR. This is why the posterior region shifts to the right in ~\ref{Fig_4_b} and \ref{Fig_4_c} as $\lambda_{\rm b}$ decreases.

The dotted lines in ~\ref{Fig_4_b} and \ref{Fig_4_c} show the $M-R$ curves for the same EoSs when $\lambda_{\rm b}$ assumes the values labelled on those figures; the solid (GR) curves are retained in these figures for reference. Both the GW170817 posteriors as well as the $M-R$ curves of the various EoSs shift to the right in the mass-radius diagram as $\lambda_{\rm b}$ decreases. However, the latter shift by a greater amount than the posteriors, which allows us to constrain $\lambda_{\rm b}$ to be greater than $35.1~{\rm GeV}^{4}$. This is because even the softest EoS in this set, namely, AP1, crosses well beyond the $M-R$ region that is allowed by the posteriors for that value of $\lambda_{\rm b}$. 

Since both $\lambda_{\rm b}$ and NS matter EoS parameters influence the GW observables $M$ and $\Lambda$, it is tricky to measure $\lambda_{\rm b}$ unless one knows the EoS in advance. For example, in GR both WFF1 and APR4 remain viable in light of GW170817 measurements. AP1 in GR is (barely) ruled out by GW170817 since it misses the GW170817 posteriors by being a little too ``soft'' (see \ref{Fig_4_a}). WFF1 remains barely viable in a Brane-world model with $\lambda_{\rm b}=44.2~{\rm GeV}^{4}$: see \ref{Fig_4_b} where the dotted yellow curve has a region of overlap with the posteriors. However, APR4 (dotted blue curve in \ref{Fig_4_b}) has no section  overlapping with the posteriors and, hence, is no longer viable in the same Brane-world model. Thus, if it were known independently, e.g., from nuclear physics experiments, that APR4 is the true EoS of NSs, then it must be the case that  $\lambda_{\rm b} >44.2~{\rm GeV}^{4}$. 

Since we do not know the true EoS, one way to constrain $\lambda_{\rm b}$ is to use the EoS that gives the smallest radius stars in GR allowed by the GW170817 posteriors. Since there are regions of the posteriors in GR that allow for smaller stars than WFF1, we instead use AP1, which bounds their radii from below and therefore, should lead to a more conservative lower bound on $\lambda_{\rm b}$. As shown in \ref{Fig_4_c}, for $\lambda_{\rm b}=35.1~{\rm GeV}^{4}$ both WFF1 and AP1 move to larger radii and are at the edge of the GW170817 posteriors. Therefore, even a slightly smaller value of $\lambda_{\rm b}$ will make both WFF1 and AP1 stars, and stars of all other EoSs shown there, disfavoured by the GW170817 posteriors. Thence, we conservatively claim that $\lambda_{\rm b} >35.1~{\rm GeV}^{4}$; see \ref{Fig_4_c}. It would be interesting to see how this constraint gets translated into different units of measurement. For example, one can use the following conversion formula connecting $\textrm{GeV}$ to $\textrm{cm}^{-1}$, which reads, $1~\textrm{GeV}=5.06 \times 10^{13}~\textrm{cm}^{-1}$. Thus the above constraint on $\lambda_{\rm b}$ gets translated to, $\lambda_{\rm b}>23.01 \times 10^{55}~\textrm{cm}^{-4}$. This can also be converted to the original unit of $\lambda_{\rm b}$, which is energy density, in which the above constraint reads, $\lambda_{\rm b}>10^{36.6}~\textrm{erg}~\textrm{cm}^{-3}$. Further, by using appropriate conversion formula, one can indeed translate this bound on $\lambda_{\rm b}$ expressed in $\textrm{GeV}^{4}$ unit to any other unit system desired.

In this context, we would also like to briefly mention another possible avenue to constrain the brane tension $\lambda_{\rm b}$, namely using the binary BH observations. The modifications brought in the post-Newtonian formalism in the presence of extra dimension will provide wealth of information from the in-spiral phase and hence may constrain $\lambda_{\rm b}$, given the fact that in presence of extra dimensions the tidal Love numbers of BHs are non-zero \cite{Chakravarti:2018vlt}. Further, the quasi-normal modes, if detected for a sufficient duration, will also constrain the brane tension $\lambda_{\rm b}$, see \cite{Chakraborty:2017qve}. These analyses are in progress and will be reported elsewhere.
\section{Concluding Remarks}\label{Sec_Conclusion}

In this work, we show how GW170817 posteriors can effectively rule out the parameter space $\lambda_{\rm b}\leq 35.1~{\rm GeV}^{4}$, which corroborates with the local bounds obtained from experiments on gravitational interaction at sub-millimetre scales~\cite{Germani:2001du,Long:2002wn}. This is a first attempt to constrain certain higher dimensional gravity model, in this case the single brane model of Randall and Sundrum. It is worth pointing out that we have found this constraint on the brane tension $\lambda_{\rm b}$ using a single binary NS merger event. Thus it is legitimate to ask, what about future observations, in particular when the upgraded LIGO, Virgo as well as KAGRA and LIGO-India joins the gravitational waves collaboration. Situation will further improve when third generation detectors, e.g., Einstein Telescope joins the quest. As we will observe more and more binary NS merger event, in future the constraint on $\lambda_{\rm b}$ will further improve. This is because, the posteriors will shrink in size and so the constraint on the brane tension will become stronger. Assuming that with $N$ observations the statistical error will shrink by a factor of $\sim\sqrt{N}$, it is easy to check that after $\sim 125$ observations, we will be able to constrain the brane tension to a value $\lambda_{b}>1.1\times 10^{3}~\textrm{GeV}^{4}$, which is comparable to the constraint available from torsion-balance experiments.

\begin{figure}[!htbp]
\centering
\includegraphics[scale=0.47]{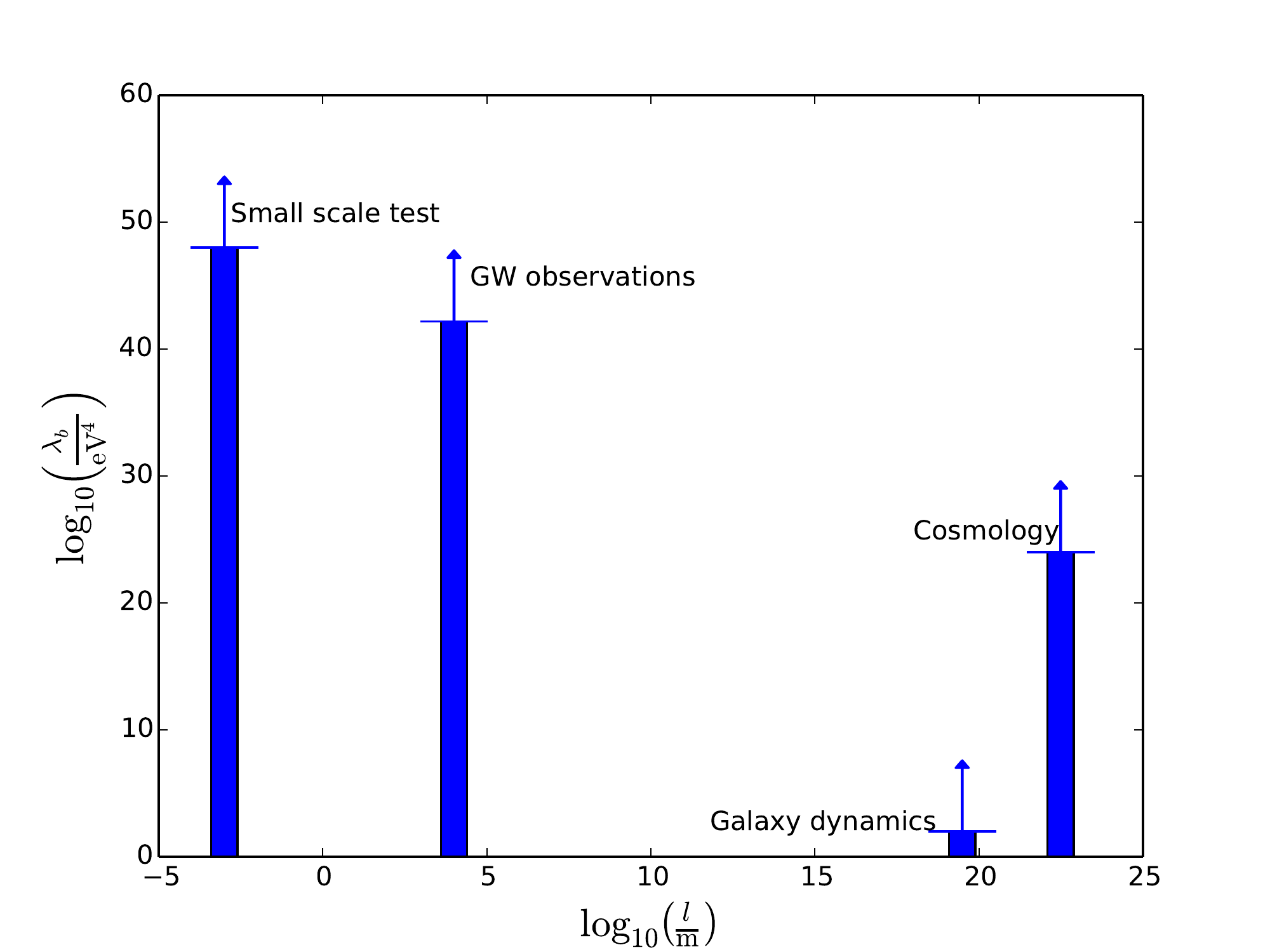}
\caption{Bounds on the brane tension $\lambda_{\rm b}$ have been depicted from various observations performed at various length scales. The bound from small scale ``table-top" experiment appears on the extreme left while from cosmology appears on the extreme right. The smallest among these bounds are from galactic dynamics, while the current bound from gravitational wave observation is presented at the second position in the above figure. Both the axes are plotted in Logarithmic scales to accommodate all the constraints appropriately. It is evident that with a single observation, the constraints arising out of the gravitational wave measurement is far more superior than those obtained from galactic or cosmological observations.}
\label{Fig_cons}
\end{figure}

Thus, with more binary NS merger observations~\cite{LIGOScientific:2018mvr,Bose:2017jvk}, and complementary information harnessed from their electromagnetic counterparts (such as tighter lower limit on $\Lambda$)~\cite{Radice:2017lry,Forbes:2019xaz}, the bound on brane tension can be  tightened further. It is worth emphasizing that the constraint on the brane tension presented here is much stronger than those derived in the context of cosmological ($\lambda_{\rm b}>1~\textrm{MeV}^{4}$) and galactic ($\lambda_{\rm b}>100~\textrm{eV}^{4}$) scales and provide the estimation of $\lambda_{\rm b}$ from an experiment on a completely different length scale (see, e.g., \ref{Fig_cons}).

Besides, a rigorous analysis of the constraint on the brane tension will require one to work with the waveform, properly tuned with the modified gravitational field equations originating from higher dimensional spacetime. However, calculating the gravitational waveforms of merging NSs in higher dimensions require significant numerical resources, which is currently beyond our scope. Still, the analysis presented here accurately encodes the nature of the modifications brought in by higher spacetime dimensions at the leading order. This can be understood by performing a quick computation of the modified phasing of the GW waveform due to the presence of $\lambda_{\rm b}$. Since the bulk Weyl stress modified the exterior geometry as well, such a change in the phasing can be explicitly computed, yielding a correction $\sim 1.5\%$ to the general relativistic prediction, which is indeed a very small contribution. Thus the modifications in the waveform due to the presence of higher dimensions will be minimal and can be ignored in the present scenario. This is true from a data analysis perspective as well, since higher order contributions in the full waveform will always be sub-dominant. Even then, the waveform modelling with higher spacetime dimensions will not only enable us to understand how the effects of higher dimensions can be embedded in gravitational waveforms, but possibly will also significantly improve the constraint on $\lambda_{\rm b}$. This will be of utmost importance, as future generations of gravitational wave detectors comes into the picture. Thus our results provide the basic landscape on which we would like to improve upon in the  future by further refining the analysis, e.g., by modelling the NS merger with appropriate wave forms.

Moreover, we would also like to emphasize that, deriving the constraint on $\lambda_{\rm b}$ is just one aspect of this work. We have also shown another very non-trivial result associated with extra dimensions, which to our knowledge has not appeared earlier and will be indispensable for any future research concerning physics of NSs in presence of higher dimensions. This has to do with the fact that universality relations among moment of inertia, tidal Love number and compactness of the NS, hold true even in the case of extra dimension, though depart from the general relativistic counterpart. It is this departure that has enabled us to provide constraint on the brane tension using gravitational wave data from GW170817 event. Our result is an explicit demonstration of the fact that the root of universality is deeper than what we have suspected in the past, since the presence of extra dimension modifies the EoS parameter non-linearly (see \ref{Eq_rho_p}). This explicitly demonstrates that possibly the fact that all NSs in the limit of infinite compaction lead to BHs is the key reason behind existence of such universality relations, which have interesting consequences regarding the no-hair theorem. Thus this work connects several, apparently, disjoint areas of physics and may lead to further interesting consequences, not only in the context of gravitational waves, but also in other areas of fundamental physics, e.g., extra dimensions and BHs.

\section*{acknowledgments}

Research of SC is supported by INSPIRE Faculty Fellowship (Reg. No. DST/INSPIRE/04/2018/000893) and he also thanks IUCAA, Pune for warm hospitality where a part of this work was done. Research of S.S.G is partially supported by the SERB-Extra Mural Research grant (EMR/2017/001372), Government of India. This work was supported in part by NSF Grant PHY-1506497 and the Navajbai Ratan Tata Trust. This document has been ascribed a LIGO document number LIGO-P1900080. The authors thank Rajesh Nayak for carefully reading the manuscript and making helpful suggestions. This research has made use of data, software and/or web tools obtained from the Gravitational Wave Open Science Center (https://www.gw-openscience.org), a service of LIGO Laboratory, the LIGO Scientific Collaboration and the Virgo Collaboration. LIGO is funded by the U.S. National Science Foundation. Virgo is funded by the French Centre National de Recherche Scientifique (CNRS), the Italian Istituto Nazionale della Fisica Nucleare (INFN) and the Dutch Nikhef, with contributions by Polish and Hungarian institutes.
\appendix
\labelformat{section}{Appendix #1} 
\labelformat{subsection}{Appendix #1}
\section{Tolman-Oppenheimer-Volkoff equations in the brane world}\label{AppA}

In this appendix, we will briefly discuss the Tolman-Oppenheimer-Volkoff equations for the braneworld scenario. These equations will govern the equilibrium structure of a NS in the braneworld paradigm and can be determined by noting that \ref{Eff_Eq} can be cast as Einstein's equations with an effective matter source involving both $\Pi_{\mu \nu}$ as well as $E_{\mu \nu}$. The energy momentum tensor and $\Pi_{\mu \nu}$ has been combined to yield the effective energy density and pressure as in \ref{Eq_rho_p}. Moreover, in the context of static and spherically symmetric spacetime the bulk Weyl tensor can be decomposed into irreducible parts, depending on an energy density $\mathcal{U}$ and pressure $\mathcal{P}$. Thus for a static and spherically symmetric metric of the following form,
\begin{align}
ds^{2}=-e^{\nu(e)}dt^{2}+e^{\lambda(r)}dr^{2}+r^{2}d\Omega^{2}~,
\end{align}
the ToV equations can be expressed as,
\begin{align}
\frac{dm}{dr}&=4\pi r^{2}\left(\rho_{\rm eff}+\frac{6\mathcal{U}}{\lambda_{\rm b}(8\pi G_{4})^{2}}\right)
\\
\frac{dp}{dr}&=-\left(\rho+p\right)\frac{\nu'}{2}
\\
\frac{\nu'}{2}&=\frac{G_{4}m+4\pi G_{4}r^{3}\left(p_{\rm eff}+\frac{2\mathcal{U}+4\mathcal{P}}{\lambda_{\rm b}(8\pi G_{4})^{2}}\right)}{r(r-2G_{4}m)}
\\
\frac{d\mathcal{U}}{dr}+\left(4\mathcal{U}+2\mathcal{P}\right)\frac{\nu'}{2}&=-2\left(4\pi G_{4}\right)^{2}\left(\rho+p\right)\frac{d\rho}{dr}-2\frac{d\mathcal{P}}{dr}-\frac{6}{r}\mathcal{P}
\end{align}
These equations, must be supplemented with appropriate boundary conditions. Here we assume that inside the star $\mathcal{U}$ and $\mathcal{P}$ are vanishing, while in the exterior $\rho$ and $p$ identically vanishes. Thus the effect of $\rho$ and $p$ in the interior is being balanced by the non-zero Weyl stress in the exterior. These equations have been solved with an appropriate EoS parameter between $\rho$ and $p$ to determine various properties of the NS. 

\bibliography{TLNReferences}

\providecommand{\href}[2]{#2}\begingroup\raggedright\begin{thebibliography}{10}

\bibitem{Abbott:2017vtc}
{\bfseries VIRGO, LIGO Scientific} Collaboration, B.~P. Abbott {\em et~al.},
  ``{GW170104: Observation of a 50-Solar-Mass Binary Black Hole Coalescence at
  Redshift 0.2},'' \href{http://dx.doi.org/10.1103/PhysRevLett.118.221101}{{\em
  Phys. Rev. Lett.} {\bfseries 118} no.~22, (2017) 221101},
\href{http://arxiv.org/abs/1706.01812}{{\ttfamily arXiv:1706.01812 [gr-qc]}}.

\bibitem{TheLIGOScientific:2016src}
{\bfseries Virgo, LIGO Scientific} Collaboration, B.~P. Abbott {\em et~al.},
  ``{Tests of general relativity with GW150914},''
  \href{http://dx.doi.org/10.1103/PhysRevLett.116.221101}{{\em Phys. Rev.
  Lett.} {\bfseries 116} no.~22, (2016) 221101},
\href{http://arxiv.org/abs/1602.03841}{{\ttfamily arXiv:1602.03841 [gr-qc]}}.

\bibitem{Abbott:2016blz}
{\bfseries Virgo, LIGO Scientific} Collaboration, B.~P. Abbott {\em et~al.},
  ``{Observation of Gravitational Waves from a Binary Black Hole Merger},''
  \href{http://dx.doi.org/10.1103/PhysRevLett.116.061102}{{\em Phys. Rev.
  Lett.} {\bfseries 116} no.~6, (2016) 061102},
\href{http://arxiv.org/abs/1602.03837}{{\ttfamily arXiv:1602.03837 [gr-qc]}}.

\bibitem{TheLIGOScientific:2017qsa}
{\bfseries Virgo, LIGO Scientific} Collaboration, B.~Abbott {\em et~al.},
  ``{GW170817: Observation of Gravitational Waves from a Binary Neutron Star
  Inspiral},'' \href{http://dx.doi.org/10.1103/PhysRevLett.119.161101}{{\em
  Phys. Rev. Lett.} {\bfseries 119} no.~16, (2017) 161101},
\href{http://arxiv.org/abs/1710.05832}{{\ttfamily arXiv:1710.05832 [gr-qc]}}.

\bibitem{Lorimer:2008se}
D.~R. Lorimer, ``{Binary and Millisecond Pulsars},''
  \href{http://dx.doi.org/10.12942/lrr-2008-8}{{\em Living Rev. Rel.}
  {\bfseries 11} (2008) 8},
\href{http://arxiv.org/abs/0811.0762}{{\ttfamily arXiv:0811.0762 [astro-ph]}}.

\bibitem{Abbott:2018lct}
{\bfseries LIGO Scientific, Virgo} Collaboration, B.~P. Abbott {\em et~al.},
  ``{Tests of General Relativity with GW170817},''
\href{http://arxiv.org/abs/1811.00364}{{\ttfamily arXiv:1811.00364 [gr-qc]}}.

\bibitem{LIGOScientific:2019fpa}
{\bfseries LIGO Scientific, Virgo} Collaboration, B.~P. Abbott {\em et~al.},
  ``{Tests of General Relativity with the Binary Black Hole Signals from the
  LIGO-Virgo Catalog GWTC-1},''
\href{http://arxiv.org/abs/1903.04467}{{\ttfamily arXiv:1903.04467 [gr-qc]}}.

\bibitem{Randall:1999ee}
L.~Randall and R.~Sundrum, ``{A Large mass hierarchy from a small extra
  dimension},'' \href{http://dx.doi.org/10.1103/PhysRevLett.83.3370}{{\em
  Phys.Rev.Lett.} {\bfseries 83} (1999) 3370--3373},
\href{http://arxiv.org/abs/hep-ph/9905221}{{\ttfamily arXiv:hep-ph/9905221
  [hep-ph]}}.

\bibitem{Randall:1999vf}
L.~Randall and R.~Sundrum, ``{An Alternative to compactification},''
  \href{http://dx.doi.org/10.1103/PhysRevLett.83.4690}{{\em Phys.Rev.Lett.}
  {\bfseries 83} (1999) 4690--4693},
\href{http://arxiv.org/abs/hep-th/9906064}{{\ttfamily arXiv:hep-th/9906064
  [hep-th]}}.

\bibitem{Chamblin:1999by}
A.~Chamblin, S.~Hawking, and H.~Reall, ``{Brane world black holes},''
  \href{http://dx.doi.org/10.1103/PhysRevD.61.065007}{{\em Phys.Rev.}
  {\bfseries D61} (2000) 065007},
\href{http://arxiv.org/abs/hep-th/9909205}{{\ttfamily arXiv:hep-th/9909205
  [hep-th]}}.

\bibitem{Shiromizu:1999wj}
T.~Shiromizu, K.-i. Maeda, and M.~Sasaki, ``{The Einstein equation on the
  3-brane world},'' \href{http://dx.doi.org/10.1103/PhysRevD.62.024012}{{\em
  Phys.Rev.} {\bfseries D62} (2000) 024012},
\href{http://arxiv.org/abs/gr-qc/9910076}{{\ttfamily arXiv:gr-qc/9910076
  [gr-qc]}}.

\bibitem{Green:1987sp}
M.~B. Green, J.~H. Schwarz, and E.~Witten, {\em {SUPERSTRING THEORY. VOL. 1:
  INTRODUCTION}}.
\newblock Cambridge Monographs on Mathematical Physics. 1988.
\newblock
\url{http://www.cambridge.org/us/academic/subjects/physics/theoretical-physics-and-mathematical-physics/superstring-theory-volume-1}.
\newblock

\bibitem{ArkaniHamed:1998rs}
N.~Arkani-Hamed, S.~Dimopoulos, and G.~Dvali, ``{The Hierarchy problem and new
  dimensions at a millimeter},''
  \href{http://dx.doi.org/10.1016/S0370-2693(98)00466-3}{{\em Phys.Lett.}
  {\bfseries B429} (1998) 263--272},
\href{http://arxiv.org/abs/hep-ph/9803315}{{\ttfamily arXiv:hep-ph/9803315
  [hep-ph]}}.

\bibitem{Antoniadis:1998ig}
I.~Antoniadis, N.~Arkani-Hamed, S.~Dimopoulos, and G.~Dvali, ``{New dimensions
  at a millimeter to a Fermi and superstrings at a TeV},''
  \href{http://dx.doi.org/10.1016/S0370-2693(98)00860-0}{{\em Phys.Lett.}
  {\bfseries B436} (1998) 257--263},
\href{http://arxiv.org/abs/hep-ph/9804398}{{\ttfamily arXiv:hep-ph/9804398
  [hep-ph]}}.

\bibitem{Rubakov:1983bb}
V.~Rubakov and M.~Shaposhnikov, ``{Do We Live Inside a Domain Wall?},''
\href{http://dx.doi.org/10.1016/0370-2693(83)91253-4}{{\em Phys.Lett.}
  {\bfseries B125} (1983) 136--138}.

\bibitem{Csaki:2004ay}
C.~Csaki, ``{TASI lectures on extra dimensions and branes},'' in {\em {From
  fields to strings: Circumnavigating theoretical physics. Ian Kogan memorial
  collection (3 volume set)}}, pp.~605--698.
\newblock 2004.
\newblock
\href{http://arxiv.org/abs/hep-ph/0404096}{{\ttfamily arXiv:hep-ph/0404096
  [hep-ph]}}.
\newblock

\bibitem{Dadhich:2000am}
N.~Dadhich, R.~Maartens, P.~Papadopoulos, and V.~Rezania, ``{Black holes on the
  brane},'' \href{http://dx.doi.org/10.1016/S0370-2693(00)00798-X}{{\em
  Phys.Lett.} {\bfseries B487} (2000) 1--6},
\href{http://arxiv.org/abs/hep-th/0003061}{{\ttfamily arXiv:hep-th/0003061
  [hep-th]}}.

\bibitem{Maartens:2003tw}
R.~Maartens, ``{Brane world gravity},''
  \href{http://dx.doi.org/10.12942/lrr-2004-7}{{\em Living Rev. Rel.}
  {\bfseries 7} (2004) 7},
\href{http://arxiv.org/abs/gr-qc/0312059}{{\ttfamily arXiv:gr-qc/0312059
  [gr-qc]}}.

\bibitem{Chandrasekhar:1985kt}
S.~Chandrasekhar, ``{The mathematical theory of black holes},'' in {\em
  {Oxford, UK: Clarendon (1992) 646 p., OXFORD, UK: CLARENDON (1985) 646 P.}}
\newblock
1985.
\newblock

\bibitem{Nollert:1999ji}
H.-P. Nollert, ``{TOPICAL REVIEW: Quasinormal modes: the characteristic `sound'
  of black holes and neutron stars},''
\href{http://dx.doi.org/10.1088/0264-9381/16/12/201}{{\em Class. Quant. Grav.}
  {\bfseries 16} (1999) R159--R216}.

\bibitem{Kokkotas:1999bd}
K.~D. Kokkotas and B.~G. Schmidt, ``{Quasinormal modes of stars and black
  holes},'' \href{http://dx.doi.org/10.12942/lrr-1999-2}{{\em Living Rev. Rel.}
  {\bfseries 2} (1999) 2},
\href{http://arxiv.org/abs/gr-qc/9909058}{{\ttfamily arXiv:gr-qc/9909058
  [gr-qc]}}.

\bibitem{Berti:2009kk}
E.~Berti, V.~Cardoso, and A.~O. Starinets, ``{Quasinormal modes of black holes
  and black branes},''
  \href{http://dx.doi.org/10.1088/0264-9381/26/16/163001}{{\em Class. Quant.
  Grav.} {\bfseries 26} (2009) 163001},
\href{http://arxiv.org/abs/0905.2975}{{\ttfamily arXiv:0905.2975 [gr-qc]}}.

\bibitem{Konoplya:2011qq}
R.~A. Konoplya and A.~Zhidenko, ``{Quasinormal modes of black holes: From
  astrophysics to string theory},''
  \href{http://dx.doi.org/10.1103/RevModPhys.83.793}{{\em Rev. Mod. Phys.}
  {\bfseries 83} (2011) 793--836},
\href{http://arxiv.org/abs/1102.4014}{{\ttfamily arXiv:1102.4014 [gr-qc]}}.

\bibitem{Toshmatov:2016bsb}
B.~Toshmatov, Z.~Stuchlik, J.~Schee, and B.~Ahmedov, ``{Quasinormal frequencies
  of black hole in the braneworld},''
  \href{http://dx.doi.org/10.1103/PhysRevD.93.124017}{{\em Phys. Rev.}
  {\bfseries D93} no.~12, (2016) 124017},
\href{http://arxiv.org/abs/1605.02058}{{\ttfamily arXiv:1605.02058 [gr-qc]}}.

\bibitem{Chakraborty:2017qve}
S.~Chakraborty, K.~Chakravarti, S.~Bose, and S.~SenGupta, ``{Signatures of
  extra dimensions in gravitational waves from black hole quasinormal modes},''
  \href{http://dx.doi.org/10.1103/PhysRevD.97.104053}{{\em Phys. Rev.}
  {\bfseries D97} no.~10, (2018) 104053},
\href{http://arxiv.org/abs/1710.05188}{{\ttfamily arXiv:1710.05188 [gr-qc]}}.

\bibitem{Chakraborty:2014xla}
S.~Chakraborty and S.~SenGupta, ``{Spherically symmetric brane spacetime with
  bulk $f(\mathcal {R})$ gravity},''
  \href{http://dx.doi.org/10.1140/epjc/s10052-014-3234-3}{{\em Eur.Phys.J.}
  {\bfseries C75} no.~1, (2015) 11},
\href{http://arxiv.org/abs/1409.4115}{{\ttfamily arXiv:1409.4115 [gr-qc]}}.

\bibitem{Chakraborty:2015taq}
S.~Chakraborty and S.~SenGupta, ``{Spherically symmetric brane in a bulk of
  f(R) and Gauss-Bonnet Gravity},''
  \href{http://dx.doi.org/10.1088/0264-9381/33/22/225001}{{\em Class. Quant.
  Grav.} {\bfseries 33} no.~22, (2016) 225001},
\href{http://arxiv.org/abs/1510.01953}{{\ttfamily arXiv:1510.01953 [gr-qc]}}.

\bibitem{Chakraborty:2015bja}
S.~Chakraborty and S.~SenGupta, ``{Effective gravitational field equations on
  $m$-brane embedded in n-dimensional bulk of Einstein and $f(\mathcal {R})$
  gravity},'' \href{http://dx.doi.org/10.1140/epjc/s10052-015-3768-z}{{\em Eur.
  Phys. J.} {\bfseries C75} no.~11, (2015) 538},
\href{http://arxiv.org/abs/1504.07519}{{\ttfamily arXiv:1504.07519 [gr-qc]}}.

\bibitem{Bhattacharya:2016naa}
S.~Bhattacharya and S.~Chakraborty, ``{Constraining some Horndeski gravity
  theories},'' \href{http://dx.doi.org/10.1103/PhysRevD.95.044037}{{\em Phys.
  Rev.} {\bfseries D95} no.~4, (2017) 044037},
\href{http://arxiv.org/abs/1607.03693}{{\ttfamily arXiv:1607.03693 [gr-qc]}}.

\bibitem{Mukherjee:2017fqz}
S.~Mukherjee and S.~Chakraborty, ``{Horndeski theories confront the Gravity
  Probe B experiment},''
  \href{http://dx.doi.org/10.1103/PhysRevD.97.124007}{{\em Phys. Rev.}
  {\bfseries D97} no.~12, (2018) 124007},
\href{http://arxiv.org/abs/1712.00562}{{\ttfamily arXiv:1712.00562 [gr-qc]}}.

\bibitem{Banerjee:2017hzw}
I.~Banerjee, S.~Chakraborty, and S.~SenGupta, ``{Excavating black hole
  continuum spectrum: Possible signatures of scalar hairs and of higher
  dimensions},'' \href{http://dx.doi.org/10.1103/PhysRevD.96.084035}{{\em Phys.
  Rev.} {\bfseries D96} no.~8, (2017) 084035},
\href{http://arxiv.org/abs/1707.04494}{{\ttfamily arXiv:1707.04494 [gr-qc]}}.

\bibitem{Lattimer:2004pg}
J.~M. Lattimer and M.~Prakash, ``{The physics of neutron stars},''
  \href{http://dx.doi.org/10.1126/science.1090720}{{\em Science} {\bfseries
  304} (2004) 536--542},
\href{http://arxiv.org/abs/astro-ph/0405262}{{\ttfamily arXiv:astro-ph/0405262
  [astro-ph]}}.

\bibitem{Hinderer:2007mb}
T.~Hinderer, ``{Tidal Love numbers of neutron stars},''
  \href{http://dx.doi.org/10.1086/533487}{{\em Astrophys. J.} {\bfseries 677}
  (2008) 1216--1220},
\href{http://arxiv.org/abs/0711.2420}{{\ttfamily arXiv:0711.2420 [astro-ph]}}.

\bibitem{Flanagan:2007ix}
E.~E. Flanagan and T.~Hinderer, ``{Constraining neutron star tidal Love numbers
  with gravitational wave detectors},''
  \href{http://dx.doi.org/10.1103/PhysRevD.77.021502}{{\em Phys. Rev.}
  {\bfseries D77} (2008) 021502},
\href{http://arxiv.org/abs/0709.1915}{{\ttfamily arXiv:0709.1915 [astro-ph]}}.

\bibitem{Binnington:2009bb}
T.~Binnington and E.~Poisson, ``{Relativistic theory of tidal Love numbers},''
  \href{http://dx.doi.org/10.1103/PhysRevD.80.084018}{{\em Phys. Rev.}
  {\bfseries D80} (2009) 084018},
\href{http://arxiv.org/abs/0906.1366}{{\ttfamily arXiv:0906.1366 [gr-qc]}}.

\bibitem{Damour:2009vw}
T.~Damour and A.~Nagar, ``{Relativistic tidal properties of neutron stars},''
  \href{http://dx.doi.org/10.1103/PhysRevD.80.084035}{{\em Phys. Rev.}
  {\bfseries D80} (2009) 084035},
\href{http://arxiv.org/abs/0906.0096}{{\ttfamily arXiv:0906.0096 [gr-qc]}}.

\bibitem{Chakravarti:2018vlt}
K.~Chakravarti, S.~Chakraborty, S.~Bose, and S.~SenGupta, ``{Tidal Love numbers
  of black holes and neutron stars in the presence of higher dimensions:
  Implications of GW170817},''
  \href{http://dx.doi.org/10.1103/PhysRevD.99.024036}{{\em Phys. Rev.}
  {\bfseries D99} no.~2, (2019) 024036},
\href{http://arxiv.org/abs/1811.11364}{{\ttfamily arXiv:1811.11364 [gr-qc]}}.

\bibitem{Cardoso:2017cfl}
V.~Cardoso, E.~Franzin, A.~Maselli, P.~Pani, and G.~Raposo, ``{Testing
  strong-field gravity with tidal Love numbers},''
  \href{http://dx.doi.org/10.1103/PhysRevD.95.089901,
  10.1103/PhysRevD.95.084014}{{\em Phys. Rev.} {\bfseries D95} no.~8, (2017)
  084014}, \href{http://arxiv.org/abs/1701.01116}{{\ttfamily arXiv:1701.01116
  [gr-qc]}}.
[Addendum: Phys. Rev.D95,no.8,089901(2017)].

\bibitem{Yazadjiev:2018xxk}
S.~S. Yazadjiev, D.~D. Doneva, and K.~D. Kokkotas, ``{Tidal Love numbers of
  neutron stars in $f(R)$ gravity},''
  \href{http://dx.doi.org/10.1140/epjc/s10052-018-6285-z}{{\em Eur. Phys. J.}
  {\bfseries C78} no.~10, (2018) 818},
\href{http://arxiv.org/abs/1803.09534}{{\ttfamily arXiv:1803.09534 [gr-qc]}}.

\bibitem{Olmo:2019flu}
G.~J. Olmo, D.~Rubiera-Garcia, and A.~Wojnar, ``{Stellar structure models in
  modified theories of gravity: lessons and challenges},''
\href{http://arxiv.org/abs/1912.05202}{{\ttfamily arXiv:1912.05202 [gr-qc]}}.

\bibitem{Yagi:2013bca}
K.~Yagi and N.~Yunes, ``{I-Love-Q},''
  \href{http://dx.doi.org/10.1126/science.1236462}{{\em Science} {\bfseries
  341} (2013) 365--368},
\href{http://arxiv.org/abs/1302.4499}{{\ttfamily arXiv:1302.4499 [gr-qc]}}.

\bibitem{Long:2002wn}
J.~C. Long, H.~W. Chan, A.~B. Churnside, E.~A. Gulbis, M.~C.~M. Varney, and
  J.~C. Price, ``{Upper limits to submillimeter-range forces from extra
  space-time dimensions},''
  \href{http://arxiv.org/abs/hep-ph/0210004}{{\ttfamily arXiv:hep-ph/0210004
  [hep-ph]}}.
[Nature421,922(2003)].

\bibitem{Germani:2001du}
C.~Germani and R.~Maartens, ``{Stars in the brane world},''
  \href{http://dx.doi.org/10.1103/PhysRevD.64.124010}{{\em Phys. Rev.}
  {\bfseries D64} (2001) 124010},
\href{http://arxiv.org/abs/hep-th/0107011}{{\ttfamily arXiv:hep-th/0107011
  [hep-th]}}.

\bibitem{Wiseman:2001xt}
T.~Wiseman, ``{Relativistic stars in Randall-Sundrum gravity},''
  \href{http://dx.doi.org/10.1103/PhysRevD.65.124007}{{\em Phys. Rev.}
  {\bfseries D65} (2002) 124007},
\href{http://arxiv.org/abs/hep-th/0111057}{{\ttfamily arXiv:hep-th/0111057
  [hep-th]}}.

\bibitem{Visser:2002vg}
M.~Visser and D.~L. Wiltshire, ``{On brane data for brane world stars},''
  \href{http://dx.doi.org/10.1103/PhysRevD.67.104004}{{\em Phys. Rev.}
  {\bfseries D67} (2003) 104004},
\href{http://arxiv.org/abs/hep-th/0212333}{{\ttfamily arXiv:hep-th/0212333
  [hep-th]}}.

\bibitem{Ovalle:2014uwa}
J.~Ovalle, L.~A. Gergely, and R.~Casadio, ``{Brane-world stars with a solid
  crust and vacuum exterior},''
  \href{http://dx.doi.org/10.1088/0264-9381/32/4/045015}{{\em Class. Quant.
  Grav.} {\bfseries 32} (2015) 045015},
\href{http://arxiv.org/abs/1405.0252}{{\ttfamily arXiv:1405.0252 [gr-qc]}}.

\bibitem{Herrera-Aguilar:2015koa}
A.~Herrera-Aguilar, A.~M. Kuerten, and R.~da~Rocha, ``{Regular Bulk Solutions
  in Brane-worlds with Inhomogeneous Dust and Generalized Dark Radiation},''
  \href{http://dx.doi.org/10.1155/2015/359268}{{\em Adv. High Energy Phys.}
  {\bfseries 2015} (2015) 359268},
\href{http://arxiv.org/abs/1501.07629}{{\ttfamily arXiv:1501.07629 [gr-qc]}}.

\bibitem{Felipe:2016lvp}
R.~González~Felipe, D.~Manreza~Paret, and A.~Pérez~Martínez, ``{Constraints
  on the braneworld from compact stars},''
  \href{http://dx.doi.org/10.1140/epjc/s10052-016-4177-7}{{\em Eur. Phys. J.}
  {\bfseries C76} no.~6, (2016) 337},
\href{http://arxiv.org/abs/1601.01973}{{\ttfamily arXiv:1601.01973 [gr-qc]}}.

\bibitem{HeydariFard:2009tj}
M.~Heydari-Fard and H.~R. Sepangi, ``{Spherically symmetric solutions and
  gravitational collapse in brane-worlds},''
  \href{http://dx.doi.org/10.1088/1475-7516/2009/02/029}{{\em JCAP} {\bfseries
  0902} (2009) 029},
\href{http://arxiv.org/abs/0903.0066}{{\ttfamily arXiv:0903.0066 [gr-qc]}}.

\bibitem{Demorest:2010bx}
P.~Demorest, T.~Pennucci, S.~Ransom, M.~Roberts, and J.~Hessels, ``{Shapiro
  Delay Measurement of A Two Solar Mass Neutron Star},''
  \href{http://dx.doi.org/10.1038/nature09466}{{\em Nature} {\bfseries 467}
  (2010) 1081--1083},
\href{http://arxiv.org/abs/1010.5788}{{\ttfamily arXiv:1010.5788
  [astro-ph.HE]}}.

\bibitem{Antoniadis:2013pzd}
J.~Antoniadis {\em et~al.}, ``{A Massive Pulsar in a Compact Relativistic
  Binary},'' \href{http://dx.doi.org/10.1126/science.1233232}{{\em Science}
  {\bfseries 340} (2013) 6131},
\href{http://arxiv.org/abs/1304.6875}{{\ttfamily arXiv:1304.6875
  [astro-ph.HE]}}.

\bibitem{Lattimer:2015nhk}
J.~M. Lattimer and M.~Prakash, ``{The Equation of State of Hot, Dense Matter
  and Neutron Stars},''
  \href{http://dx.doi.org/10.1016/j.physrep.2015.12.005}{{\em Phys. Rept.}
  {\bfseries 621} (2016) 127--164},
\href{http://arxiv.org/abs/1512.07820}{{\ttfamily arXiv:1512.07820
  [astro-ph.SR]}}.

\bibitem{Linares:2018ppq}
M.~Linares, T.~Shahbaz, and J.~Casares, ``{Peering into the dark side:
  Magnesium lines establish a massive neutron star in PSR J2215+5135},''
  \href{http://dx.doi.org/10.3847/1538-4357/aabde6}{{\em Astrophys. J.}
  {\bfseries 859} no.~1, (2018) 54},
\href{http://arxiv.org/abs/1805.08799}{{\ttfamily arXiv:1805.08799
  [astro-ph.HE]}}.

\bibitem{Cromartie:2019kug}
H.~T. Cromartie {\em et~al.}, ``{Relativistic Shapiro delay measurements of an
  extremely massive millisecond pulsar},''
  \href{http://dx.doi.org/10.1038/s41550-019-0880-2}{{\em Nat. Astron.}
  {\bfseries 4} no.~1, (2019) 72--76},
\href{http://arxiv.org/abs/1904.06759}{{\ttfamily arXiv:1904.06759
  [astro-ph.HE]}}.

\bibitem{Yagi:2013awa}
K.~Yagi and N.~Yunes, ``{I-Love-Q Relations in Neutron Stars and their
  Applications to Astrophysics, Gravitational Waves and Fundamental Physics},''
  \href{http://dx.doi.org/10.1103/PhysRevD.88.023009}{{\em Phys. Rev.}
  {\bfseries D88} no.~2, (2013) 023009},
\href{http://arxiv.org/abs/1303.1528}{{\ttfamily arXiv:1303.1528 [gr-qc]}}.

\bibitem{Yagi:2016bkt}
K.~Yagi and N.~Yunes, ``{Approximate Universal Relations for Neutron Stars and
  Quark Stars},'' \href{http://dx.doi.org/10.1016/j.physrep.2017.03.002}{{\em
  Phys. Rept.} {\bfseries 681} (2017) 1--72},
\href{http://arxiv.org/abs/1608.02582}{{\ttfamily arXiv:1608.02582 [gr-qc]}}.

\bibitem{Gupta:2017vsl}
T.~Gupta, B.~Majumder, K.~Yagi, and N.~Yunes, ``{I-Love-Q Relations for Neutron
  Stars in dynamical Chern Simons Gravity},''
  \href{http://dx.doi.org/10.1088/1361-6382/aa9c68}{{\em Class. Quant. Grav.}
  {\bfseries 35} no.~2, (2018) 025009},
\href{http://arxiv.org/abs/1710.07862}{{\ttfamily arXiv:1710.07862 [gr-qc]}}.

\bibitem{Vallisneri:2014vxa}
M.~Vallisneri, J.~Kanner, R.~Williams, A.~Weinstein, and B.~Stephens, ``{The
  LIGO Open Science Center},''
  \href{http://dx.doi.org/10.1088/1742-6596/610/1/012021}{{\em J. Phys. Conf.
  Ser.} {\bfseries 610} no.~1, (2015) 012021},
\href{http://arxiv.org/abs/1410.4839}{{\ttfamily arXiv:1410.4839 [gr-qc]}}.

\bibitem{Abbott:2018exr}
{\bfseries Virgo, LIGO Scientific} Collaboration, B.~P. Abbott {\em et~al.},
  ``{GW170817: Measurements of neutron star radii and equation of state},''
  \href{http://dx.doi.org/10.1103/PhysRevLett.121.161101}{{\em Phys. Rev.
  Lett.} {\bfseries 121} no.~16, (2018) 161101},
\href{http://arxiv.org/abs/1805.11581}{{\ttfamily arXiv:1805.11581 [gr-qc]}}.

\bibitem{LIGOScientific:2018mvr}
{\bfseries LIGO Scientific, Virgo} Collaboration, B.~P. Abbott {\em et~al.},
  ``{GWTC-1: A Gravitational-Wave Transient Catalog of Compact Binary Mergers
  Observed by LIGO and Virgo during the First and Second Observing Runs},''
\href{http://arxiv.org/abs/1811.12907}{{\ttfamily arXiv:1811.12907
  [astro-ph.HE]}}.

\bibitem{Bose:2017jvk}
S.~Bose, K.~Chakravarti, L.~Rezzolla, B.~S. Sathyaprakash, and K.~Takami,
  ``{Neutron-star Radius from a Population of Binary Neutron Star Mergers},''
  \href{http://dx.doi.org/10.1103/PhysRevLett.120.031102}{{\em Phys. Rev.
  Lett.} {\bfseries 120} no.~3, (2018) 031102},
\href{http://arxiv.org/abs/1705.10850}{{\ttfamily arXiv:1705.10850 [gr-qc]}}.

\bibitem{Radice:2017lry}
D.~Radice, A.~Perego, F.~Zappa, and S.~Bernuzzi, ``{GW170817: Joint Constraint
  on the Neutron Star Equation of State from Multimessenger Observations},''
  \href{http://dx.doi.org/10.3847/2041-8213/aaa402}{{\em Astrophys. J.}
  {\bfseries 852} no.~2, (2018) L29},
\href{http://arxiv.org/abs/1711.03647}{{\ttfamily arXiv:1711.03647
  [astro-ph.HE]}}.

\bibitem{Forbes:2019xaz}
M.~M. Forbes, S.~Bose, S.~Reddy, D.~Zhou, A.~Mukherjee, and S.~De,
  ``{Constraining the neutron-matter equation of state with gravitational
  waves},''
\href{http://arxiv.org/abs/1904.04233}{{\ttfamily arXiv:1904.04233
  [astro-ph.HE]}}.

\end{thebibliography}\endgroup

\bibliographystyle{./utphys1}

\end{document}